\documentclass[aps,prb,twocolumn,showpacs, notitlepage]{revtex4-2}
\usepackage{graphicx}
\usepackage{natbib}
\usepackage[utf8]{inputenc}
\usepackage{epstopdf}
\setcitestyle{super}
\usepackage{color}
\usepackage{amsmath}
\usepackage[normalem]{ulem}
\usepackage{pdfpages}
\usepackage{hyperref}
\makeatletter
\AtBeginDocument{\let\LS@rot\@undefined}
\makeatother

\usepackage{xcolor}

\usepackage{hyphenat}
\begin{document}

\preprint{APS/123-QED}

\title{Tuning superconductivity and spin-vortex fluctuations in CaKFe$_4$As$_4$ through in-plane antisymmetric strains}

\author{Adrian Valadkhani$^1$}
\author{Belén Zúñiga Céspedes$^2$}
\author{Salony Mandloi$^2$}
\author{Mingyu Xu$^{3,4}$}
\author{Juan Schmidt$^{3,4}$}
\author{Sergey L. Bud'ko$^{3,4}$}
\author{Paul C. Canfield$^{3,4}$}
\author{Roser Valent\'i$^{1}$}
\author{Elena Gati$^{2}$}\email{elena.gati@cpfs.mpg.de}

\address{$^{1}$ Institute for Theoretical Physics, Goethe University Frankfurt, 60438 Frankfurt am Main, Germany}
\address{$^{2}$ Max Planck Institute for Chemical Physics of Solids, 01187 Dresden, Germany}
\address{$^{3}$ Ames National Laboratory, US Department of Energy, Iowa State University, Ames, Iowa 50011, USA}
\address{$^{4}$ Department of Physics and Astronomy, Iowa State University, Ames, Iowa 50011, USA}

\date{\today}

\begin{abstract}

Lattice strains of appropriate symmetry have served as an excellent tool to explore the interaction of superconductivity in the iron-based superconductors with nematic and stripe spin-density wave (SSDW) order, which are both closely tied to an orthorhombic distortion. In this work, we contribute to a broader understanding of the coupling of strain to superconductivity and competing normal-state orders by studying CaKFe$_4$As$_4$ under large, in-plane strains of $B_{1g}$ and $B_{2g}$ symmetry. In contrast to the majority of iron-based superconductors, pure CaKFe$_4$As$_4$ exhibits superconductivity with relatively high transition temperature of $T_c\,\sim\,$35\,K in proximity of a non-collinear, tetragonal, hedgehog spin-vortex crystal (SVC) order. Through experiments, we demonstrate an anisotropic in-plane strain response of $T_c$, which is reminiscent of the behavior of other pnictides with nematicity. However, our calculations suggest that in CaKFe$_4$As$_4$, this anisotropic response correlates with the one of the SVC fluctuations, highlighting the close interrelation of magnetism and high-$T_c$ superconductivity. By suggesting moderate $B_{2g}$ strains as an effective parameter to change the stability of SVC and SSDW, we outline a pathway to a unified phase diagram of iron-based superconductivity.

\end{abstract}

\pacs{xxx}

\maketitle

The phase diagrams of high-temperature superconductors typically show various competing ordering tendencies, the fluctuations of which might be considered as the main pairing glue for superconductivity \cite{Fradkin15}. It is often found that the competing electronic orders are accompanied by the formation of a pronounced in-plane anisotropy \cite{Fradkin10,Fernandes14,Hinkov08}. This observation has initiated a huge surge in using external stresses and strains of appropriate symmetry, that couple directly to the anisotropic electronic state \cite{Hicks14,Kim20,Chu12,Ikeda18,Willa19,Gati20,Boehmer22}, as  a tool to explore the role of its fluctuations for superconductivity.

\begin{figure}
    \centering
    \includegraphics[width = \columnwidth]{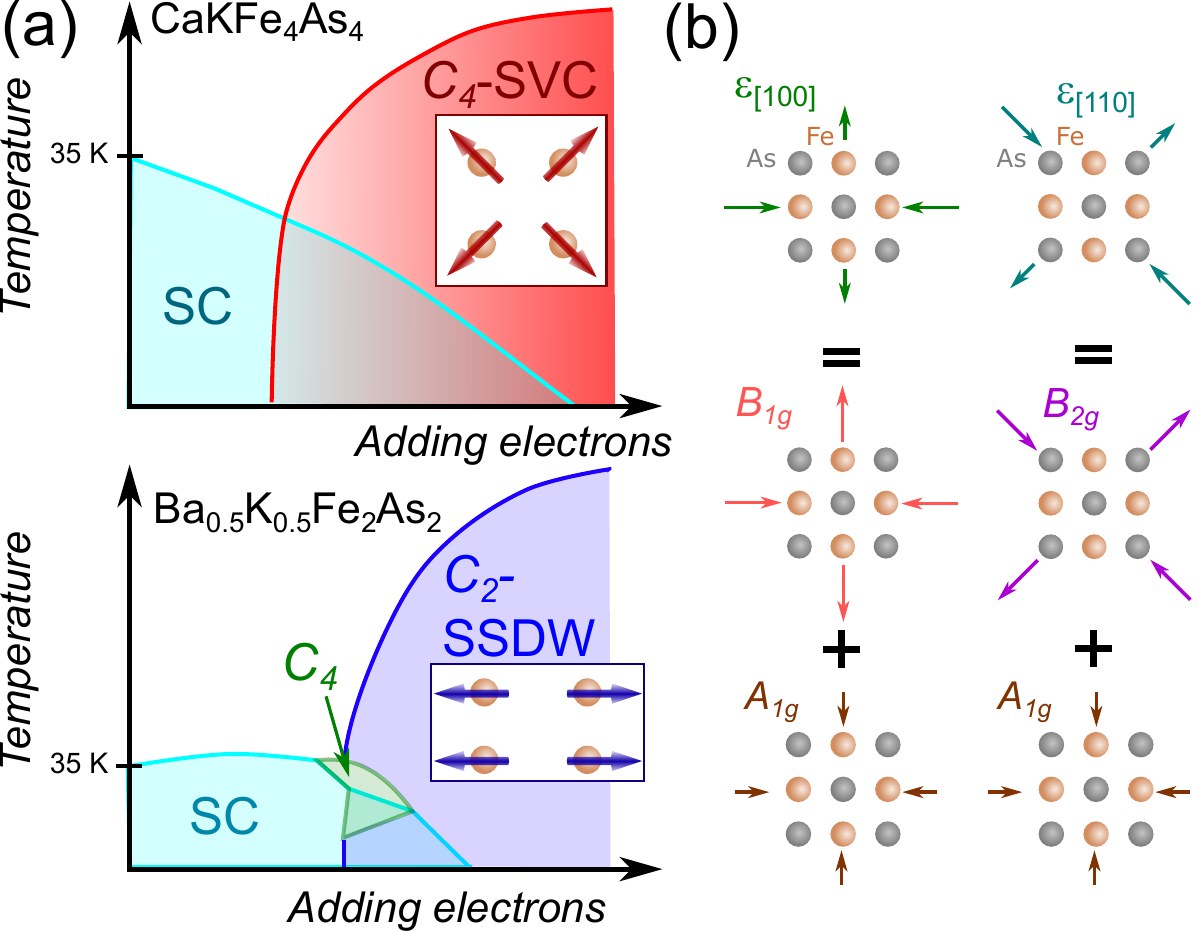}
    \caption{(a) Schematic temperature-doping phase diagrams of the two superconductors CaKFe$_4$As$_4$ ('1144', after Ref. \cite{Meier18}) and Ba$_{0.5}$K$_{0.5}$Fe$_2$As$_2$ ('122', after Ref. \cite{Boehmer15}).  Upon adding electrons to the systems, superconductivity (SC, light blue) is suppressed and magnetic phases emerge. For electron-doped 1144, the magnetic phase is the so-called spin-vortex (SVC) phase that preserves the tetragonal $C_4$-symmetry. In contrast, for the 122 compounds, the magnetic phase is the stripe spin-density wave order (SSDW), which displays $C_2$ symmetry and is accompanied by a vestigial nematic phase. Only in proximity of both SC and $C_2$-SSDW order, a small region of $C_4$ magnetic order can be observed in the 122 compound; (b) Schematic representation of the symmetry decomposition of applied strains with respect to the tetragonal unit cell. $A_{1g}$ strains preserve the tetragonal symmetry, whereas $B_{1g}$ and $B_{2g}$ do not (after Ref.~\cite{Ikeda18}). The induced $B_{1g}$ and $B_{2g}$ strains are larger than the $A_{1g}$ strains due to the in-plane Poisson ratio, see SI.}
    \label{fig:overview}
\end{figure}

In this context, iron-based superconductors have served as prime examples to explore and establish the intimate connection between anisotropic electronic states and superconductivity. Many of the members of this family, such as $Ae_{1-x}A_x$Fe$_2$As$_2$ ($Ae$\,=\,Ba, Sr, Ca and $A$\,=\,K, Na) or Ba(Fe$_{1-x}T_x$)$_2$As$_2$ ($T$\,=\,Co, Rh, Ni, Pd) \cite{Canfield10} with 1-2-2 stochiometry, show superconductivity in proximity of stripe-type spin-density wave (SSDW) magnetism\cite{Dai15} (see Fig.\,\ref{fig:overview}\,(a)). The SSDW order is characterized by ordering vectors $\textbf{Q}_\textrm{SSDW} = (\pi,0)$ or $(0,\pi)$, which gives rise to an inequivalence between the two in-plane directions of the high-temperature tetragonal lattice. This type of magnetism therefore results, aside from broken spin-rotational symmetry, in a spontaneous $B_{2g}$ lattice distortion, which reduces the crystallographic symmetry from $C_4$ to $C_2$. The order with broken lattice symmetry but preserved time-reversal symmetry is commonly referred to as nematic order. It is therefore often found to be 'vestigial' to the SSDW order \cite{Fernandes13,Fernandes19}. In other cases, like FeSe \cite{Boehmer17}, the nematic phase is even more prominent, as it is not accompanied by the formation of long-range SSDW order at ambient pressure.

The understanding of normal-state of iron-based superconductors has been tremendously advanced by utilizing lattice strains of different symmetry \cite{Ikeda18,Willa19,Gati20,Boehmer22}. This was, for example, crucial in establishing the electronically-driven nature of nematicity \cite{Chu12,Fernandes14,Worasaran21,Boehmer22}. In order to provide compelling evidence that superconductivity benefits from this unusual, anisotropic normal state, a set of experiments recently studied the direct impact of applied lattice strains on the superconducting critical temperature $T_c$ in a series of tetragonal 122 compounds \cite{Malinowski20,Liu19a}. They found that, whereas $B_{2g}$ strains, which break the same symmetry as the nematic order, measurably suppress $T_c$ both under compression and tension, the application of strain in the $B_{1g}$ channel (i.e., a strain that is oriented 45$^\circ$ away from the nematic axis) has resulted in a much weaker response \cite{Liu19a}. Based on a phenomenological Landau model, the anisotropic strain-response of $T_c$ was attributed to the coupling thereof to the nematic order parameter. This has strengthened the notion that nematic fluctuations play a key role\cite{Lederer15,Boehmer22} in boosting $T_c$.

However, this notion might be questioned \cite{Zhang18,Ding18} by the discovery of superconductivity with very high $T_c$ values in proximity to magnetic and charge-ordered states that preserve the tetragonal $C_4$ symmetry. In this context, a particular notable reference material is the quarternary compound CaKFe$_4$As$_4$, which is a superconductor with high $T_c\,\approx\,$35\,K \cite{Iyo16,Meier16} in its stochiometric form, i.e., free from substitutional disorder. This superconducting state occurs in proximity of a $C_4$ magnetic state \cite{Meier18,Kreyssig18,Budko18,Kaluarachchi17,Xiang18,Boehmer20,Xu22,Xu23} (see Fig.\,\ref{fig:overview}\,(a)). In this so-called spin-vortex type (SVC) magnetic order, the moments rotate clock-wise/anti-clock-wise in an Fe plaquette \cite{Fernandes16} ($\textbf{Q}_\textrm{SVC} = (\pi,\pi)$). Importantly, upon tuning by doping and hydrostatic pressure \cite{Meier18,Kaluarachchi17,Xiang18,Xu22}, so far no $C_2$ symmetric order has been identified. This phenomenology has motivated proposals that isotropic magnetic fluctuations, related to the SVC order, are sufficient to generate high-temperature superconductivity \cite{Zhang18,Ding18}.

Clearly, CaKFe$_4$As$_4$ presents a unique and possibly much richer platform to explore the coupling of superconductivity and its normal-state properties to in-plane strains, compared to the majority of iron-based superconductors with prominent SSDW magnetism and nematicity (see Fig.\,\ref{fig:overview}\,(a)). Yet, the evolution of the superconductivity and magnetism in CaKFe$_4$As$_4$ have not been studied under large, tunable in-plane strains, even though such studies promise key insights for developing a unified understanding of the phase diagram of iron-based superconductors.

In this work, we combine experiments and density-functional theory (DFT) calculations to shed light on this issue. We first demonstrate that the superconducting $T_c$ of CaKFe$_4$As$_4$ shows a strong anisotropic response under antisymmetric in-plane strains of $B_{1g}$ and $B_{2g}$ symmetry, reminiscent of the findings in 122 compounds. We then show through calculations, that in-plane strains in CaKFe$_4$As$_4$ have primarily a different effect than in 122 compounds. In CaKFe$_4$As$_4$, moderate, experimentally achievable, in-plane strains of $B_{2g}$ symmetry offer an excellent mean to change the preference from an SVC to SSDW state. In contrast, $B_{1g}$ strains leave the preference for SVC unchanged. Taken together, these results suggest that the development of in-plane anisotropic strain response is not a unique fingerprint of coupling to nematicity, but can also arise from the coupling of superconductivity to fluctuations of the non-collinear SVC magnetic order. With these results, we provide important insights to a broader understanding of strain tuning of the multiple phases in iron-based superconductors.

For clarity, we will use the notion of irreducible strains throughout this paper. For the tetragonal unit cell of CaKFe$_4$As$_4$, there are two antisymmetric irreducible strains, denoted by $\epsilon_{B1g}$ and $\epsilon_{B2g}$, which break $C_4$ symmetry. In this work, $B_{1g}$ and $B_{2g}$ refer to the irreducible representations of the tetragonal point group associated with the actual crystallographic unit cell rather than the one-Fe unit cell. As schematically shown in Fig.\,\ref{fig:overview}\,(b), $\epsilon_{B1g}$ and $\epsilon_{B2g}$ strains are primarily induced when strain is applied along the crystallographic [1\,0\,0] and [1\,1\,0] directions, respectively, in addition to a smaller fully symmetric strain $\epsilon_{A1g}$. The used decomposition procedure is described in the SI \cite{SI}. Tensile (compressive) strains are denoted throughout our work by positive (negative) signs.

We first demonstrate how the superconducting critical temperature, $T_c$, changes with these in-plane antisymmetric strains through experiments. To this end, oriented CaKFe$_4$As$_4$ crystals\cite{Meier17} (along [1\,0\,0] and [1\,1\,0]) were mounted on rigid platforms\cite{Park20,Bartlett21}. Varying strains were applied to the platform and the sample with a piezoelectric-actuator-based uniaxial pressure cell \cite{Hicks14}. $T_c$ was determined through temperature-dependent measurements of the mutual inductance of two concentric coils wound around the platform with the sample (see SI Sec. \ref{sec:methods-SI} for details).

\begin{figure}
    \centering
    \includegraphics[width = 1\columnwidth]{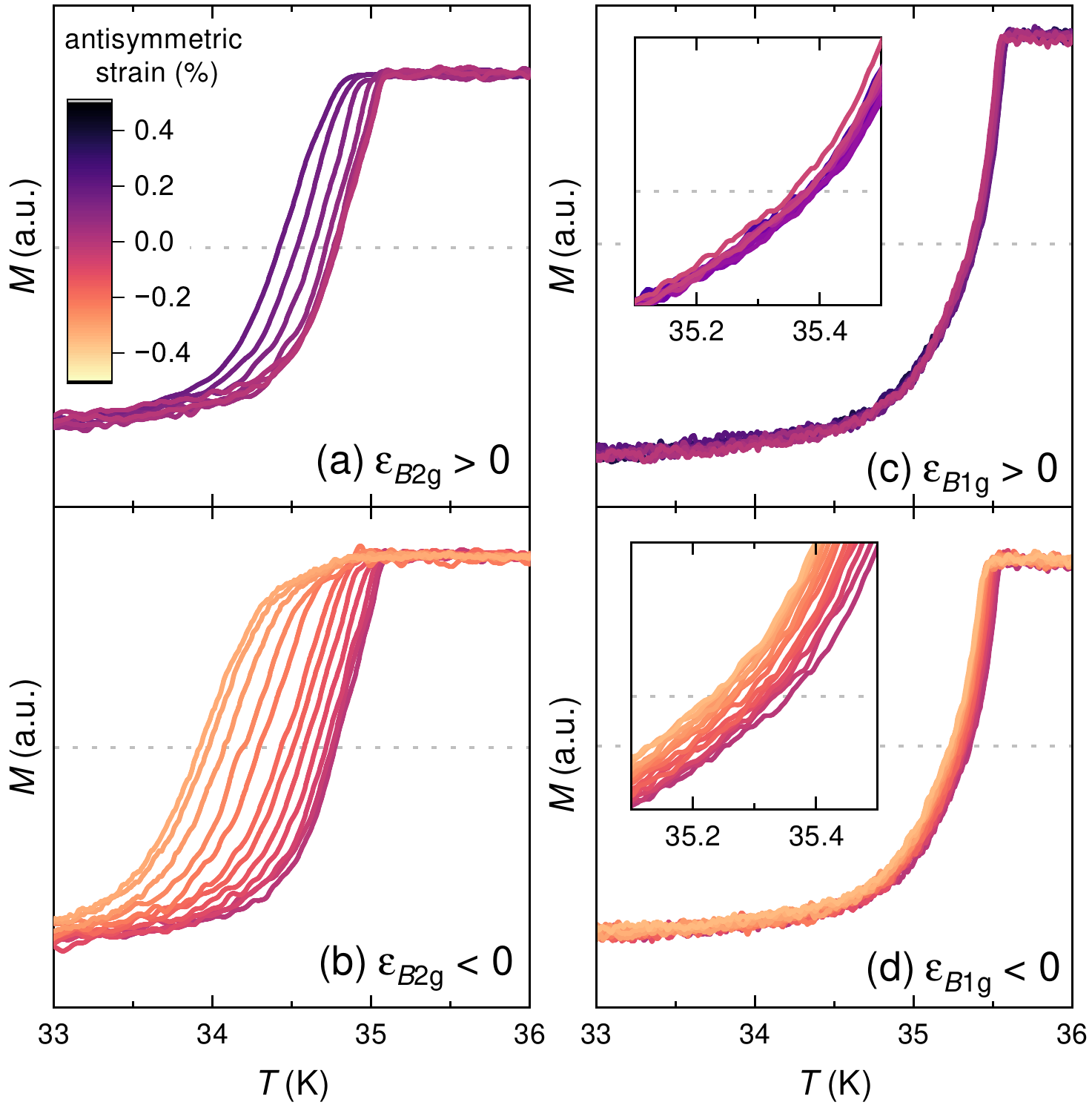}
    \caption{Mutual inductance data on CaKFe$_4$As$_4$ as a function of temperature for different applied, antisymmetric strains  (a)-(b) $\epsilon_{B2g}$ and (c)-(d) $\epsilon_{B1g}$. Top (bottom) panels show data under tensile (compressive) strains. The spacing in strain between two data sets amounts to $\sim\,0.05\%$. The grey dashed line visualizes the 50\% threshold that is used to infer $T_c$.}
    \label{fig:mutual-inductance}
\end{figure}

Figure \ref{fig:mutual-inductance} shows the temperature dependence of the mutual inductance, $M$, at different strains of type (a)-(b) $\epsilon_{B2g}$  and (c)-(d) $\epsilon_{B1g}$. In each figure, the top (bottom) panel shows the data taken under tensile (compressive) strains. The superconducting transition is clearly identified in all data sets by a sharp drop of $M$ and we assign $T_c$ to the temperature where $M$ has reached 50\% of its full value (see grey dashed line). The bare $M$ data reveals clearly our main experimental findings. First, the response of $T_c$ to $\epsilon_{B2g}$ is larger than the one to $\epsilon_{B1g}$ strain (Note the same scale of the temperature axes in all plots). Second, both compressive and tensile $\epsilon_{B2g}$ strains suppress $T_c$. The suppression is as large as $\Delta T_c\,\sim\,$-0.8\,K by $\epsilon_{B2g}\sim\,-\,0.4\%$.

\begin{figure}
    \centering
    \includegraphics[width = \columnwidth]{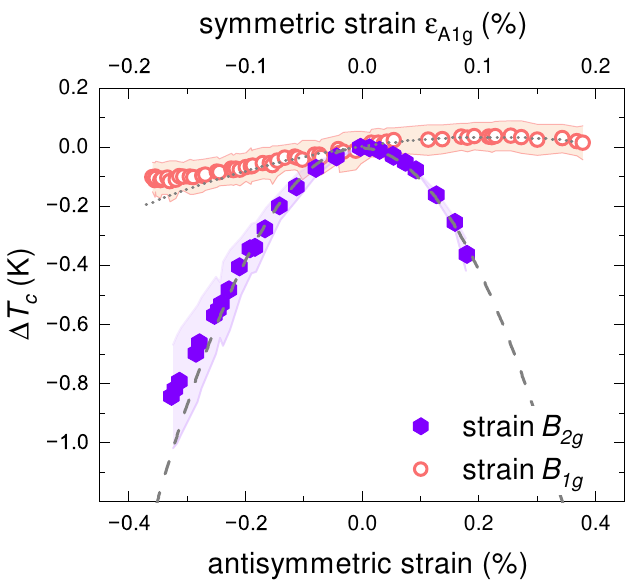}
    \caption{Change of superconducting critical temperature of CaKFe$_4$As$_4$, $\Delta T_c\,=\,T_c(\epsilon)-T_c(\epsilon\,=\,0)$ with antisymmetric $\epsilon_{B2g}$ (closed symbols) and $\epsilon_{B1g}$ (open symbols) strains (bottom axis). In both experiments under dominating $B_{1g}$ and $B_{2g}$ strains, a finite symmetric $A_{1g}$ strain is induced as well and depicted on the top axis. The color shading represents the width of the transition. Dashed and dotted grey lines correspond to polynomial fits up to the second order in strain.}
    \label{fig:Tc-vs-strain}
\end{figure}

To further quantify the statements above, we compiled the phase diagram as $\Delta T_c\,=\,T_c (\epsilon)-T_c (\epsilon\,=\,0)$ up to $\,\pm\,0.4\%$ antisymmetric strains (and $\,\pm\,0.2\%$ symmetric strains) in Fig.\,\ref{fig:Tc-vs-strain}. The color shaded areas around the $T_c (\epsilon)$ data indicate the width of the superconducting transition at each $\epsilon$, determined from 25\% and 75\% threshold values. 

By consideration of symmetry-allowed terms in a Ginzburg-Landau approach \cite{Malinowski20,Fernandes13}, it is expected that, to lowest order, $\Delta T_c (\epsilon)\, \sim D_{A1g} \epsilon_{A1g} +  D_{i} \epsilon_{i}^2$ with $i=B_{1g}$ or $B_{2g}$. The linear strain dependence can only result from the dependence of $T_c$ on $\epsilon_{A1g}$, which are also induced in our experiments.

Figure \ref{fig:Tc-vs-strain} shows that the data of $T_c$ vs. $\epsilon_{B2g}$ is clearly dominated by the quadratic strain dependence, expected for antisymmetric strains, over almost the full strain range. A polynomial fit of order two (see dashed line) yields the quadratic coefficient $D_{B2g}/T_c(0)\,=\,-(2780\,\pm\,50)\,$. Only for high compression ($|\epsilon|\gtrsim\,0.3\%$), small deviations from the quadratic behavior are observed, which, however, are still within the error bars of our experiment.

For the $B_{1g}$ data, a weak quadratic change of $T_c$ with $\epsilon_{B1g}$ can also be identified, even though the linear contribution to $T_c(\epsilon)$ due to $A_{1g}$ strains dominates. The quadratic coefficient amounts to $D_{B1g}/T_c(0)\,=\,-(36\,\pm\,1)\,$, which is unlikely to result from a small misalignment of the crystal (see SI) and is therefore considered intrinsic to the $B_{1g}$ channel. Similar to the $B_{2g}$ data, only small deviations from the polynomial fit within the error bars of the experiment occur for $|\epsilon|\gtrsim\,0.3\%$.



\begin{figure}
    \centering
    \includegraphics[width = 0.8\columnwidth]{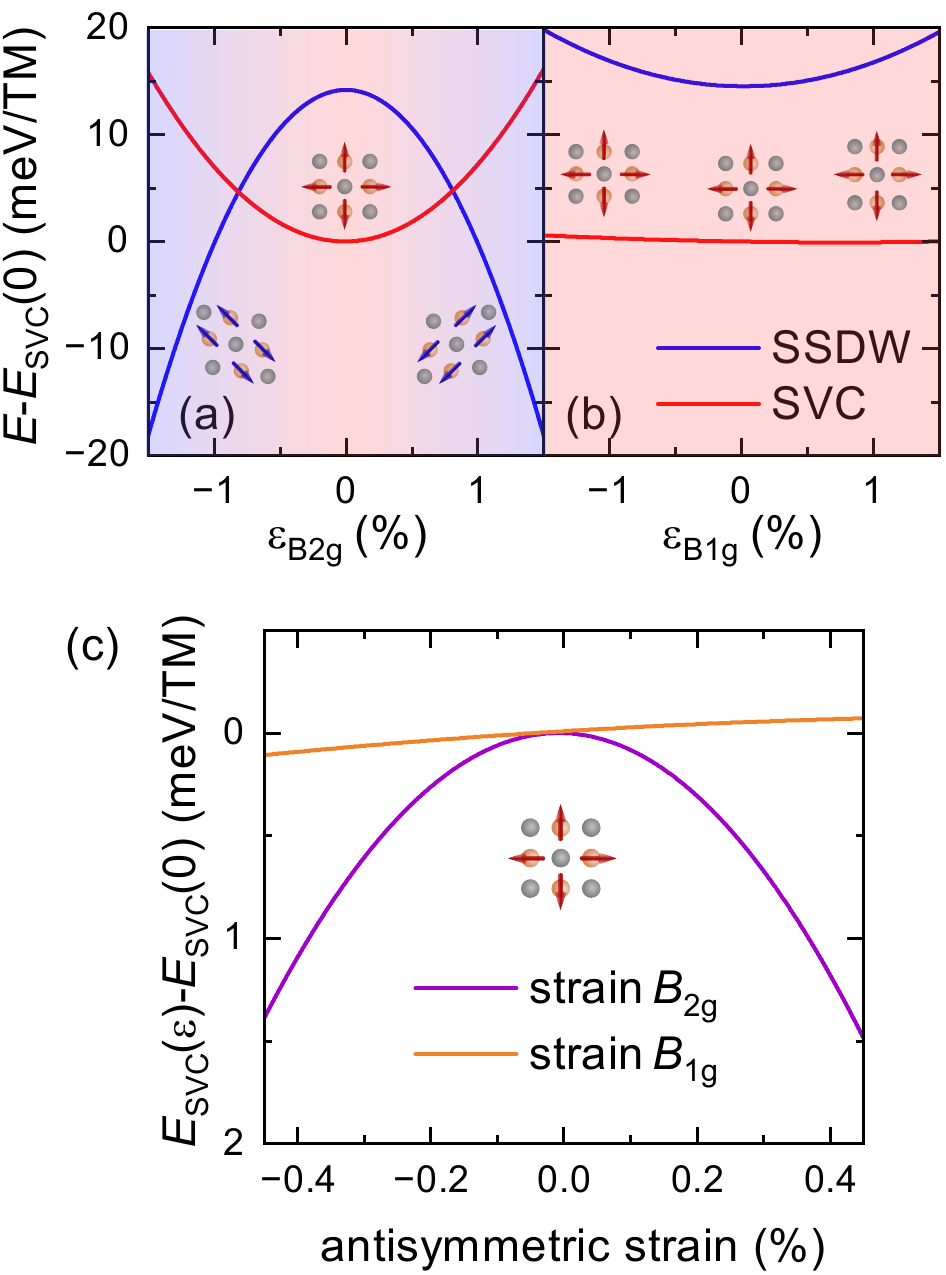}
    \caption{(a,b) Calculated energies of CaKFe$_4$As$_4$ for imposed ''frozen'' spin configurations of spin-vortex crystal (SVC) order and stripe spin-density wave (SSDW) order as a function of antisymmetric strains of  (a) $B_{2g}$ and (b) $B_{1g}$  symmetry. Whereas for $B_{1g}$ strains the SVC configuration remains clearly energetically favorable, $B_{2g}$ strains change the preferred type of spin fluctuations from SVC to SSDW around $\epsilon_{B2g}\,\approx\,\pm\,0.8\,\%$, i.e., when $E_\textrm{SSDW}(\epsilon)<E_\textrm{SVC}(\epsilon)$. The SVC (SSDW) ordering motif is visualized by red (blue) arrows in the small cartoons; (c) Energy of the SVC state as a function of antisymmetric strains on enlarged scales in the experimentally studied strain range.}
    \label{fig:calculations-magorder}
\end{figure}

To explore possible correlations of $T_c(\epsilon)$ with the strain dependence of the normal-state fluctuations, we discuss in the following our results of DFT calculations under the same antisymmetric strain fields. In CaKFe$_4$As$_4$ at ambient conditions, no static magnetic order \cite{Ding18,Budko17} can be found, but strong SVC fluctuations \cite{Ding18,Xie18} exist. In previous computational studies, it has been shown that it is crucial to take spin fluctuations into account for accurate predictions of the ambient-pressure structure and structural transitions at high pressures \cite{Kaluarachchi17,Borisov18,Song19}. In these works, the presence of spin fluctuations was simulated by imposing a ''frozen'' magnetic configuration within a reduced Stoner theory, in which the size of magnetic moments is adjusted for the values found in experiment in Ni-doped CaKFe$_4$As$_4$ \cite{Kreyssig18}. Given that this  approach has proven successful in exploring the coupling of magnetism to strain, we now calculate within DFT the energy of ''frozen'' SVC and SSDW orders in CaKFe$_4$As$_4$ under finite in-plane strains (see SI) and use it as a proxy for the nature and strength of magnetic fluctuations.

Consistent with earlier DFT results \cite{Kaluarachchi17,Borisov18}, the result at $\epsilon\,=\,0$ is that a SVC configuration is energetically favored over the SSDW ($E_\textrm{SVC}(0)<E_\textrm{SSDW}(0)$), see Fig.\,\ref{fig:calculations-magorder}. For finite strains, the change in energy of a given magnetic configuration is, to lowest order, given by $\Delta E\,\sim C_{A1g} \epsilon_{A1g}+C_{i}\epsilon_i^2$ with $i=B_{1g},B_{2g}$. The calculated energies for both SVC and SSDW magnetic configurations are well-described by such a linear plus quadratic strain dependence for $\epsilon_{B2g}$ (Fig.~\ref{fig:calculations-magorder} (a)) and $\epsilon_{B1g}$ (Fig.~\ref{fig:calculations-magorder} (b)). The sign and strength of the quadratic term of the strain dependence, however, strongly differ between the different orders and strains, as we discuss below. 

Specifically, $\epsilon_{B2g}$ strongly weakens the tendency towards the SVC configuration and promotes the one towards SSDW order (Fig.\,\ref{fig:calculations-magorder} (a)), consistent with a symmetry analysis within a Landau approach \cite{Meier18b}, since the SSDW couples directly to $\epsilon_{B2g}$ (see also Fig.\,\ref{fig:overview}). In contrast, the energy of the SVC configuration is only weakly increased by $\epsilon_{B1g}$. At the same time, the SSDW configuration becomes significantly unfavorable under increasing $\epsilon_{B1g}$ (Fig.\,\ref{fig:calculations-magorder} (b)). 

On a quantitative level, following important conclusions can be drawn. First, a $B_{2g}$ strain of $\epsilon_{B2g}^{crit}\approx\,\pm\,0.8\%$ changes the preferred type of spin fluctuations from SVC and SSDW. We note that $\epsilon_{B2g}^{crit}$ is larger than typical strains that are induced by a spontaneous nematic/SSDW distortion in the 122 pnictides ($\epsilon_{B2g}\,\lesssim\,0.3\%$) \cite{Boehmer15}. This reflects the fact that the CaKFe$_4$As$_4$ is not as close to a SSDW-nematic instability at ambient conditions as the related 122 compounds \cite{Boehmer20}. Second, for $B_{1g}$ strains, there is hardly any change in the magnetism and, the SVC configuration remains clearly favorable. Overall, this results in a clear anisotropy of the antisymmetric strain dependence of the magnetic energies of the SVC order of $(C_{B2g}/C_{B1g})_\textrm{SVC}\sim\,52$ (Fig. \ref{fig:calculations-magorder}\,(c)). 


The qualitatively similar anisotropic strain response of the SVC configuration in Fig.\,\ref{fig:calculations-magorder}\,(c) and the experimentally-determined superconducting $T_c$ in Fig.\,\ref{fig:Tc-vs-strain} under both types of antisymmetric strains, $\epsilon_{B1g}$ and $\epsilon_{B2g}$, is striking and is the main result of the present work. Even the magnitude of the measured anisotropy parameters of $T_c$ of $D_{B2g}/D_{B1g}\,\sim\,$77 and of $E$ of $C_{B2g}/C_{B1g}\sim\,52$ is similar. Even though the calculated energies can only serve as a rough proxy for the spin fluctuations, it is natural to assume that in a scenario of magnetically-driven superconductivity \cite{Chubukov12,Hirschfeld16,Liu22}, there exists a correlation of $E$ and $T_c$ (and correspondingly of $C_i$ and $D_i$). As a consequence, in the case of CaKFe$_4$As$_4$ with SVC configuration, it is the fact that $\epsilon_{B2g}$ strains primarily weaken SVC fluctuations \cite{Meier18b}, while $\epsilon_{B1g}$ do not, that is most likely at the origin of the observed in-plane strain anisotropy of $T_c$. Whereas the anisotropic strain response of $T_c$ is a widely observed feature of iron-based superconductivity \cite{Malinowski20,Liu19a,Zhao23}, it is driven in CaKFe$_4$As$_4$ by the coupling of superconductivity to magnetism, rather than by nematicity, as suggested for the 122 compounds. This conclusion strengthens the view that it is the magnetism that dominates the properties of high-$T_c$ iron-based superconductivity. The observation that bulk FeSe does not show a clear quadratic contribution \cite{Bartlett21,Ghini21} to $T_c(\epsilon_{B2g})$ further points to magnetic degrees of freedom \cite{Zhao23} being an important ingredient for the development of a strain-anisotropy of $T_c$.

At the same time, our results clearly demonstrate a route towards a unified phase diagram of iron-based superconductivity by using antisymmetric strains in the CaKFe$_4$As$_4$ family, since these strains might be used to manipulate the relative importance of SVC and SSDW magnetism \cite{Boehmer20} for superconductivity. The theoretical prediction of the strain tunability of magnetism (see Fig.\,\ref{fig:calculations-magorder}) within an experimentally-achievable strain range motivates a series of studies in the future. For example, it would be very interesting to study superconducting properties at larger $B_{2g}$ strains, in particular at those strains, where SSDW fluctuations become dominant. A simple extrapolation of the present $T_c(\epsilon_{B2g})$ data, using the quadratic fit function, to $\epsilon_{B2g}^{crit}\,\pm\,\sim\,0.8\%$ would predict a quite sizable $T_c$ of $\sim\,$27\,K. If such a relatively high $T_c$ can be confirmed and combined with microscopic insights \cite{Khasanov18,Jost18,Lochner17,Bristow20,Mou16,Cho17,Teknowijoyo18,Iida17,Biswas17}, the idea that both SVC and SSDW fluctuations promote similar superconducting states \cite{Fernandes16} with high $T_c$ might be strengthened. 

In summary, we have established how antisymmetric strains can be used to tune the superconductivity and the nature of magnetic fluctuations in the stochiometric high-$T_c$ superconductor CaKFe$_4$As$_4$. Specifically, we demonstrated that both superconductivity and the preferred magnetic configuration develop a highly anisotropic strain response to in-plane antisymmetric strains of $B_{2g}$ and $B_{1g}$ type. The correlation of these two quantities strongly suggests that the anisotropic response is driven by the coupling of the non-collinear SVC magnetic configuration to antisymmetric strain, rather than by nematicity. Thus, our work contributes to a broader understanding of how antisymmetric strains impact superconductivity and its competing states. Based on the prediction that moderate antisymmetric strains can be used to manipulate the relative stability of non-collinear and collinear orders, antisymmetric strain tuning is expected to be a powerful tuning parameter for a wide range of magnetic quantum materials.


\textit{Acknowledgements - }We thank William Meier and Andreas Kreyssig for insightful discussions on the 1144 compounds, as well as Jack Barraclough for useful discussions on the limitations of the strain platforms. Financial support by the Max Planck Society is gratefully acknowledged. In addition, we gratefully acknowledge funding through the Deutsche Forschungsgemeinschaft (DFG, German Research Foundation) through TRR 288—422213477 (project B05) and the SFB 1143 (project-id 247310070; project C09). Research in Dresden benefits from the environment provided by the DFG Cluster of Excellence ct.qmat (EXC 2147, project ID 390858940). Work at the Ames National Laboratory was supported by the U.S. Department of Energy, Office of Science, Basic Energy Sciences, Materials Sciences and Engineering Division. The Ames National Laboratory is operated for the U.S. Department of Energy by Iowa State University under Contract No. DEAC02-07CH11358.

\bibliographystyle{apsrev}

\clearpage

\appendix
\section{Supplemental Information}

\subsection{Extended Methods}\label{sec:methods-SI}

\textit{Experiment -} Samples of CaKFe$_4$As$_4$ were grown following the procedure described in Ref. \cite{Meier17}. For measurements under varying strains, we used a piezoelectric-actuator based uniaxial pressure cell \cite{Hicks14}. Since CaKFe$_4$As$_4$ is very malleable and cleaves easily, we relied on the method of attaching thin samples to rigid platforms that are stressed to apply large, tunable strains to mechanically delicate samples, initially described in detail in Ref.\,\cite{Park20,Bartlett21}. This method ensures as homogeneous strain as possible by avoiding buckling and minimizes the risk for cleavage of the sample. 

\begin{figure}[h]
    \centering
    \includegraphics[width = 0.7\columnwidth]{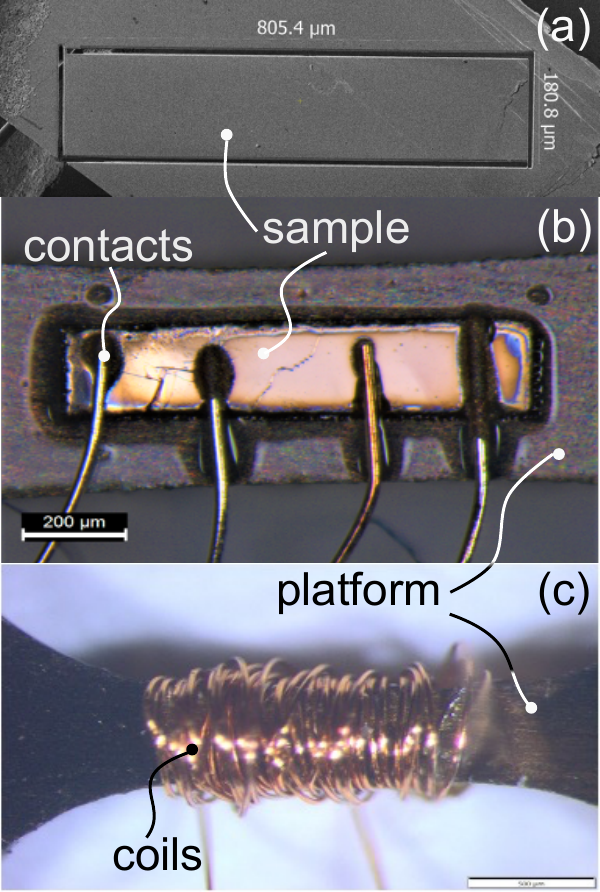}
    \caption{Experimental setup for the determination of the superconducting transition temperature of CaKFe$_4$As$_4$ under tunable in-plane strains; (a) Bar-shaped sample cut out of a larger piece of single crystal, using Plasma Focus Ion Beam milling; (b) Sample mounted on rigid platform, and contacted for four-probe resistance measurements; (c) Two coils wound around the section of the platform, on which the sample is mounted, for measurements of the mutual inductance.}
    \label{fig:exp-setup}
\end{figure}

To ensure good strain transmission into the bulk of the sample (see Ref. \cite{Park20} for finite-element simulations), the sample dimensions were typically chosen to be $\sim\,800\,\times\,180\,\times\,10\,\mu$m$^3$, with longest axis either along the [1\,0\,0] or the [1\,1\,0] direction. The in-plane dimensions and orientation were precisely controlled using Plasma Focused Ion Beam (PFIB) cutting (see Fig.\,\ref{fig:exp-setup}\,(a)). The small out-of-plane thickness was achieved by multiple cleaving steps of the samples prior to PFIB cutting. After cleaving and cutting, the samples were glued on the platforms using Stycast 1266 (see Fig.\,\ref{fig:exp-setup}\,(b)). Using this procedure, the maximum misalignment of the sample in the plane amounts to $\sim$\,3$^\circ$.

The maximum applied strain that was achieved in our experiments was $\sim\,\pm\,0.6\,\%$, corresponding to $\epsilon_{B2g}\sim\,\pm\,0.4\,\%$ (see below). This value is limited by plastic deformation of the platform material (grade 5 Ti-V-Al in our work), but not by the elastic limit\cite{Song19} of CaKFe$_4$As$_4$, which showed fully elastic behavior across the entire strain range studied and reproducible $T_c$ behavior upon various strain sweeps. We note that the elastic limit of the platform material here exceeds the report of Refs.\,\cite{Park20,Bartlett21} by a factor of 2. This was achieved by work hardening the platform inside a force-displacement cell, similar to the one described in Ref. \cite{Barber19}, at room temperature prior to mounting the sample. Monitoring both force and displacement through capacitive sensors installed in the cell (see Ref.\,\cite{Barber19} for a description of the working principle of the sensors) allows for the detection of plastic deformation of the platform. For all measurements reported here, the platform showed linear elastic behavior to a good approximation, i.e., showed a linear force-displacement relation. Only very close to the maximal strains applied, some small deviations from perfect linearity were observed. 


For the reference value $\epsilon_0\,=\,0$, we chose the strain at which $T_c$ is maximal for [1\,1\,0]$_T$ strain and used the same capacitance reading of the displacement sensor as the strain-neutral point for the data with strain along [1\,0\,0]$_T$.

The superconducting transition temperature, $T_c$, was determined through temperature-dependent measurements of the mutual inductance of two concentric coils wound around the platform with the sample (see Fig.\,\ref{fig:exp-setup}\,(c)). We used about 20 turns per coil, which was sufficient to measure the superconducting signal with high-enough signal-to-noise ratio.

In total, 10 samples were measured (8 with strain applied along the crystallographic [1\,1\,0] orientation and 2 with strain applied along the [1\,0\,0] direction) using two different platform materials (grade 5-Titanium and quartz). Results were found to be consistent among different samples and platforms, so that we focus here on results from samples mounted on the Ti platforms only.

\textit{Theory -} We perform electronic structure calculations within density functional theory (DFT)\cite{Hohenberg64,Kohn65} by using the pseudo-potential augmented plane-wave (PAW) \cite{Bloechl94,Kresse99} Vienna Ab initio Simulation Package (VASP) \cite{Kresse93,Kresse96,Kresse96_2}.
All calculations were performed with the Perdew–Burke–Ernzerhof (PBE) generalized gradient approximation (GGA) \cite{Perdew96} and include spin orbit coupling. 
The energy cut-off was set to $600$\,eV. 
For the relaxations we used a $6$\,$\times$\,$6$\,$\times$\,$4$ gamma centered $k$-mesh with a force threshold of $0.001$\,eV/\AA{} and unrestricted DFT magnetic moments.
We increased the number of $k$-points to $10$\,$\times$\,$10$\,$\times$\,$4 $ and set a convergence criterion of $10^{-6}$\,eV for the self-consistent-field
(SCF) total energies.
For all of the calculations we follow the protocol established in previous works \cite{Borisov18,Kaluarachchi17,Meier18,Boehmer20,Song19}.
In order to get information about the most stable underlying magnetic order which is responsible for the nature of the
spin fluctuations at a given strain value,  we compared the total energies of the two competing orders in this structure -- antiferromagnetic stripe and vortex "hedgehog" \cite{Borisov18, Meier18} -- to the zero-strain energy difference of the vortex and stripe state. 
The strains were applied in our calculations on a $\sqrt{2}\,\times\,\sqrt{2}\,\times\,1$ unit cell for a strain range from $-0.5\,\%$ to $0.5\,\%$ in steps of $0.25\,\%$ in each direction for each order with its unrestricted DFT magnetic moments.
The DFT ground state for the vortex order converges to a magnetic moment $\mu_{Fe,v}=1.35\mu_B$ per iron site, while the stripe order 
converges to a value of $\mu_{Fe,s}\,=\,1.47\,\mu_B$ per iron site. 
With these DFT magnetic moments $\mu_{Fe,v/s}$ the stripe order is lower in energy compared to the vortex order.
Experimental observations, however, suggest the vortex fluctuations to be the most dominant with a much lower magnetic moment $\mu_{Fe}\,\approx\,0.4\,\mu_B$ per iron site \cite{Kreyssig18}.
By carefully decreasing the size of our DFT magnetic moments in the SCF calcuations but keeping their direction and the structure of the unrestricted DFT calculation, each order has been restricted to ten different magnetic moments ranging from $0.2\,\mu_B$ to $1.5\,\mu_B$.
At about $\mu_{Fe}\,\approx0.8\,\mu_B$ and less the vortex order is lower in energy than the stripe order.
To obtain the energy $E$ of an order for a given magnetic moment $\mu_i$ we interpolated the 10 points with cubic splines and took the respective energy $E_i\,=\,E(\mu_i)$.
For the vortex order, the two strain directions are $[100]$ and $[110]$.
For the stripe order, the $[100]$ direction is treated in the same way, however, for the $[110]$ case it is important to distinguish between applying strain along or perpendicular to the stripes.
This was tested explicitely by initializing the stripe order once along $[110]$ and once along $[1\bar{1}0]$.
Of the two stripe orientation datasets, the datapoints lowest in energy were used to obtain the final fit.
The fits were determined using a second order polynomial.
For all of our fits we took the experimental reference value of $\mu_i=0.4\mu_B$~\cite{Kreyssig18} for Ni-doped CaKFe$_4$As$_4$ at zero strain to obtain the total energy in dependence of strain. 
Every order has been compared to the zero strain vortex case.
Due to the restriction of the magnetic moments to $\mu_i$ on a structure with very different DFT magnetic moment $\mu_{Fe,v/s}$ (magneto-elastic) stresses were induced to the unit cell.
Within the range of our strains, the stress tensor remains unchanged except for the component we want to change, therefore, we were able to correct the unwanted stresses by adding a linear correction term to the fitted total energies.\\

 

    \label{fig:mag-anisotropy}

\subsection{Decompositions of strains into irreducible representations}
\label{sec:irrep-strains-SI}

Here, we outline how the applied strains, $\epsilon_{[1\,0\,0]}$ and $\epsilon_{[1\,1\,0]}$, can be decomposed into irreducible representations \cite{Ikeda18}, see Fig.\,\ref{fig:overview}\,(b) for a schematic representation. The decomposition follows from the character table of point group $4/mmm$ \cite{bradley2010mathematical}.

The irreducible strains are given by following decomposition rules: $\epsilon_{A1g}\,=\,\frac{1}{2}(\epsilon_{[1\,0\,0]}+\epsilon_{[0\,1\,0]})=\,\frac{1}{2}(\epsilon_{[1\,1\,0]}+\epsilon_{[1\,\Bar{1}\,0]})$, $\epsilon_{B1g}\,=\,\frac{1}{2}(\epsilon_{[1\,0\,0]}-\epsilon_{[0\,1\,0]})$ and $\epsilon_{B2g}\,=\,\frac{1}{2}(\epsilon_{[1\,1\,0]}-\epsilon_{[1\,\Bar{1}\,0]})$. The strains along the different crystallographic directions are related by the Poisson ratio $\nu$.

In our experimental configuration, where thin samples are attached to a rigid platform, $\nu$ is given by the Poisson's ratio of the platform. Thus, for a titanium-alloy platform,  $\nu\,=\,-\epsilon_{[0\,1\,0]}/\epsilon_{[1\,0\,0]}=-\epsilon_{[1\,\Bar{1}\,0]}/\epsilon_{[1\,1\,0]}\,\sim\,0.32$ \cite{Bartlett21}. Consequently, $\epsilon_{A1g} = \frac{1}{2} (1-\nu) \epsilon_{[1\,0\,0]}=\,0.34\epsilon_{[1\,0\,0]} =\,0.34\epsilon_{[1\,1\,0]}$, $\epsilon_{B1g} =\,0.66\epsilon_{[1\,0\,0]}$ and $\epsilon_{B2g} =\,0.66\epsilon_{[1\,1\,0]}$, which has been used to decompose the data shown in the main text.

For the theoretical calculations, we use the DFT-calculated strain tensor and the same decomposition rules to determine $\epsilon_{B1g}$ and $\epsilon_{B2g}$.

 \subsection{Influence of possible small misalignments of the crystallographic axes on analysis}

In the following, we discuss how a possible, small misalignment of the crystal axes with respect to the strain axis influences our measurement results of $T_c(\epsilon)$. In the present study, this analysis is particularly relevant for measurements under $\epsilon_{B1g}$, where both $T_c(\epsilon_{B1g})$ and $T_c(\epsilon_{B2g})$ exhibit a quadratic suppression of $T_c$ with $D_{B1g}\,\ll\,D_{B2g}$ and thus, the observed $T_c(\epsilon_{B1g})$ might be significantly influenced by a small misalignment.

\begin{figure}
    \centering
    \includegraphics[width = 0.8\columnwidth]{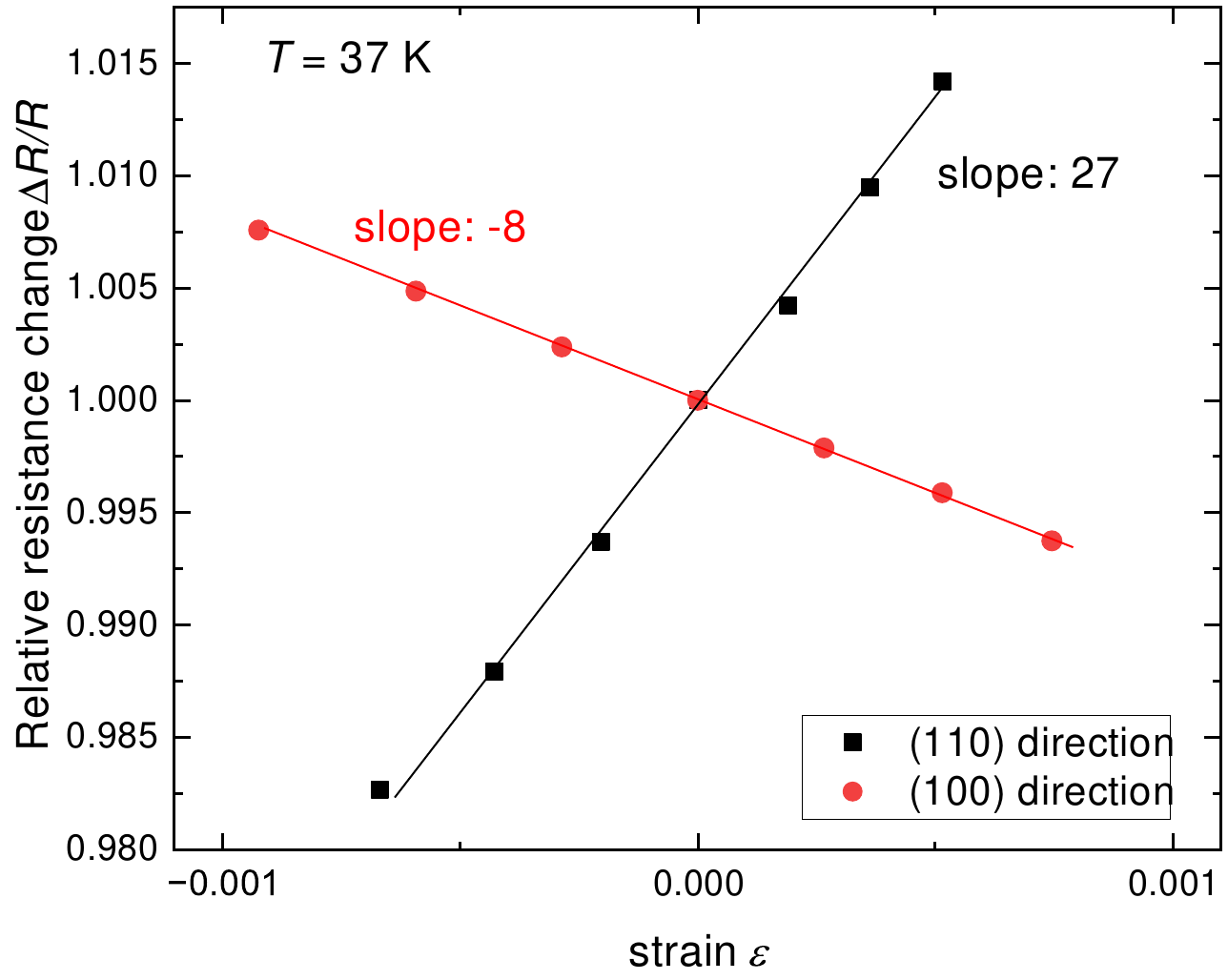}
    \caption{Longitudinal elastoresistance of CaKFe$_4$As$_4$, $\Delta R/R$, measured at $T\,=\,37\,$K\,$>\,T_c$ for strain applied along the tetragonal [110] direction (black squares) and the [100] direction (red circles).}
    \label{fig:elastoresistance}
\end{figure}

As a first confirmation of the good sample aligment, we show in Fig.\,\ref{fig:elastoresistance} the elastoresistance of the normal-state, measured at $T\,=\,37\,$K, i.e., above $T_c$. In general in many iron-based superconductors, the normal-state elastoresistance is signficant and shows a strongly anisotropic behavior, depending on whether strain is applied along the [1\,0\,0] or the [1\,1\,0] direction. This is also the case for CaKFe$_4$As$_4$, as shown in previous works \cite{Meier18,Boehmer20}, where it was shown that the longitudinal elastoresistance $m=1/R \Delta R/\Delta \epsilon$ shows a different sign for $\epsilon_{100}$ vs. $\epsilon_{110}$ and $m_{110}\sim-3m_{100}$. Our data of the elastoresistance, shown in Fig.\,\ref{fig:elastoresistance}, is fully in line with previous results, suggesting only a very minor, possible misalignment error.

For an explicit error analysis, we evaluated the possibile scenario that $D_{B1g}=0$ and the observed quadratic component in $T_c(\epsilon_{B1g})$ therefore solely results from a misalignment of the crystal, so that the contribution of $B_{2g}$ strains to the experimentally applied strain is significant. The result of this calculation showed that this scenario is only possible, if the misalignment of the crystal is around $6-7^\circ$, which is quite a bit larger than realistic error of maximally $3^\circ$. We therefore conclude that a weak quadratic suppression of $T_c$ with $\epsilon_{B1g}$ is intrinsic to CaKFe$_4$As$_4$.

\clearpage

\end{document}